\documentclass[aps,preprint,groupedaddress,showpacs]{revtex4}
\begin{document}

\title{Raman scattering study of phonon and spin modes in (TMTSF)$_2$PF$_6$}
\author{Z. V. Popovi\'c $^{a}$, V. A. Ivanov $^{b}$, O. P. Khoung $^{a}$,
, T. Nakamura $^{c}$, G. Saito $^{d}$ and  V.V. Moshchalkov
$^{a}$}

\affiliation{ $^a$ Laboratorium voor Vaste-Stoffysica en
Magnetisme, K. U. Leuven, Celestijnenlaan 200D, B-3001 Leuven,
Belgium } \affiliation{$^b$ Departement Natuurkunde, Universiteit
Antwerpen UIA, Universiteitsplein 1, B-2610 Antwerpen, Belgium }
\affiliation{$^c$ Research Institute for Electronic Science
Hokudai University, Sapporo, 060-0812 Japan} \affiliation{$^d$
Department of Chemistry, Graduate School of Science, Kyoto
University, Kyoto, 606-8501 Japan}

\begin{abstract}
We have studied polarized Raman spectra of (TMTSF)$_2$PF$_6$
single crystals in the wide spectral and temperature range using
different laser energies. We observed and assigned 23 Raman active
modes. The carbon C=C in-phase and out-of-phase stretching modes
at 1464 cm$^{-1}$ and 1602 cm$^{-1}$ show strong
electron-molecular-vibration coupling. The mode at about 1565
cm$^{-1}$ is temperature independent below 50 K due to the methyl
group motion freezing. For (TMTSF)$_2$PF$_6$ in the spin density
wave phase for a temperature induced dimensionality 1D (quarter
filled dimerized linear chain with correlated electrons) the spin
gapless magnon spectra have been calculated. No evidence of
spin-lattice coupling or spin-related modes is found in the Raman
spectra in the spin density wave phase.

KEYWORDS: Organic superconductors, Raman spectroscopy,
(TMTSF)$_2$PF$_6$, strongly correlated electrons, magnon
dispersion.
\end{abstract}

\pacs{ 78.30.Jw; 74.70.Kn; 75.30.Fv; 63.20.Dj; }  \maketitle

\section {Introduction}
The (TMTSF)$_2$PF$_6$
[bis-(tetramethyltetraselenafulvalene)hexafluorophosphate] salt is
classified as a metal since it has metal-like conductivity almost
in the whole temperature region except at very low temperatures.
Due to the instability of the quasi-one-dimensional Fermi surface,
at ambient pressure (TMTSF)$_2$PF$_6$ undergoes a phase transition
to a spin density wave (SDW) state below T$_{SDW}\simeq$12 K. The
SDW nesting is suppressed by applying pressure and this material
becomes superconducting with T$_c$=1.1 K, at p=6.5 kbar \cite{a1}.
Physical properties of this salt are discussed in details in Refs.
\cite{a2, a3}.

The (TMTSF)$_2$PF$_6$ compound has a triclinic crystal structure
\cite{a4} with $P\overline{1}$ space group and the unit cell
parameters a=7.297 \AA, b=7.711 \AA, c=13.522 \AA,
$\alpha$=83.39$^{0}$, $\beta$=86.27$^{0}$, $\gamma$=71.01$^{0}$,
Z=1. Schematic representation of crystal structure in (ac) and
(ab) plane is given in Figs. 1.(a) and 1.(b), respectively. As it
can be seen from Fig. 1, the (TMTSF)$_2$PF$_6$ crystal structure
is dominated by stacks of the TMTSF molecules arranged in a zigzag
pattern nearly perpendicular to the {\bf a}-axis \cite {a4}. The
TMTSF molecules appear to be weakly dimerized by their actual
position in the unit cell. By lowering the temperature the
triclinic stabilization becomes even stronger, as stated in
Ref.\cite {a5}.

In spite of the intense research efforts the superconducting state
in (TMTSF)$_2$PF$_6$ and other isostructural (TMTSF)$_2$X (X=
monovalent anion like ClO$_4^-$, FSO$_3^-$, PF$_6^-$ or AsF$_6^-$)
Bechgaard salts is not yet fully understood. It should be also
mentioned that these materials have attracted more attention in
the last years because of the normal state, {\it i.e.} that of a
quasi-one-dimensional metal, and because of the SDW state \cite
{a6,a7,a8,a9,a10,a11}. Below T$_{SDW}$ the presence of
antiferomagnetic ordering is confirmed by both ESR \cite{a12} and
NMR \cite{a13} measurements. The spin-density-wave gap of
(TMTSF)$_2$PF$_6$ at about $\Delta$=4.5 meV (35 cm$^{-1}$) is
obtained from the far-infrared reflectivity measurements using the
E$||$b' polarized light \cite{a9}. The electron tunneling
spectroscopy revealed even smaller value of the SDW gap: 2.5-2.9
meV \cite{a14}, which is consistent with the one obtained by the
infrared absorption, $\Delta$=2.5-4.3 meV\cite{a15}.

Phonon properties of (TMTSF)$_2$PF$_6$ were poorly studied because
the phonons are masked by the free carriers (plasmons) in the
infrared and the far-infrared spectral regions. On the other hand,
an extremely low scattering efficiency and temperature instability
of (TMTSF)$_2$PF$_6$ hindered Raman scattering measurements.
Because of that, only the vibrational properties of the TMTSF
neutral molecule and its radical cation were studied in more
details \cite {a16} by the infrared and Raman spectroscopies and
the normal coordinate analysis based on a modified valence force
field (VFF) model. The Raman scattering in organic superconductors
is not widely used because of a very weak intensity of the Raman
signal. The most studied organic superconductors by the Raman
spectroscopy are $\kappa$-(BEDF-TTF)$_2$X salts \cite
{a17,a18,a19}. To the best of our knowledge the Raman spectra of
(TMTSF)$_2$PF$_6$ were measured only in Ref.\cite {a20,a21}.
Iwahana {\it et al.} found two well shaped modes at 1463 cm$^{-1}$
and 1559 cm$^{-1}$ (see Fig. 2 of Ref. \cite{a20}). All other
modes had the intensity comparable with the level of noise.
Krauman {\it et al.} shown that the intensity of the Raman modes
of (TMTSF)$_2$PF$_6$ is strongly enhanced with 647.1 nm line of Kr
laser \cite {a21}.

In this paper we present the polarized Raman spectra of the
(TMTSF)$_2$PF$_6$ single crystals measured with different laser
excitation energies in the wide spectral and temperature range
($4K<T<300K$). Totally 23 Raman active modes are observed,
including 6 reported for the first time. The C=C in-phase and
out-of-phase stretching modes at 1464 and 1602 cm$^{-1}$ show an
equal frequency shift from both the neutral (TMTSF)$^0$ molecule
and its radical cation (TMTSF)$^+$ as a consequence of quarter
filling. The mode at about 1565 cm$^{-1}$, which represents the
CH$_3$ methyl group motion, shows no frequency shift at
temperatures below 50 K because of vibrations freezing of CH$_3$
molecule. We calculated the magnon dispersion for the dimerized
quarter-filled antiferromagnetic linear chains and did not find
the spin-gapped state in the SDW phase. The Raman spectra in the
SDW state ($4K<T<12K$) shown no evidence of spin-related features.

\section{Experimental details}
The single crystals of the (TMTSF)$_2$PF$_6$ have been prepared by
the electrocrystallization method \cite{a22}. The samples have a
needle-shaped form with typical dimensions 2 mm x 0.5 mm x 0.05 mm
along the {\bf a}, {\bf b'} and {\bf c'}- axis, respectively. The
Raman spectra were recorded in the backscattering configuration
using micro- Raman system with Dilor triple monochromator
including liquid nitrogen cooled CCD-detector. For low-temperature
measurements we used Oxford Microstat$^{He}$ continuous flow
cryostat with 0.5 mm thick window. Focusing of the laser beam was
realized with long distance (10 mm focal length) microscope
objective (magnification 50X). As an excitation source we used an
Ar-ion laser. Overheating of the samples was observed with the
laser power levels as low as 10 mW (0.6 mW at the sample). We
found that laser power level of 0.02 mW on the sample is
sufficient to obtain Raman signal and, except for the
signal-to-noise rations, no changes of the spectra were observed
as a consequence of laser heating by further lowering of laser
power. Corresponding excitation power density was less than 0.1
kW/cm$^2$.

\section{Experimental results}

Figure 2 shows the (aa) polarized low temperature (T=10 K) Raman
spectra of (TMTSF)$_2$PF$_6$ in the 1300 - 1700 cm$^{-1}$ spectral
region, excited with 457.9 nm (a) and 514.5 nm (b) energies. The
main difference between these spectra is an intensity enhancement
of 1565 cm$^{-1}$ (1602 cm$^{-1}$) mode with 457.9 nm (514.5 nm)
excitation, as a result of the resonance effect. Figure 3 shows
(b'b') polarized Raman spectrum of (TMTSF)$_2$PF$_6$ measured at
10 K in the spectral range from 150 to 1700 cm$^{-1}$. The same
spectra in the 2700-3000 cm$^{-1}$ range is given in the inset of
Fig. 3. Further lowering of temperature up to our experimental
limit of 4 K does not produce any change in the Raman spectra. As
we expected, we have got better signal to noise ratio for (b'b')
polarization (b'=insulating axis) than in the (aa) case. We
observed 23 Raman active modes in four spectral ranges: (i) 2800 -
3000 cm$^{-1}$; (ii) 1450 - 1650 cm$^{-1}$; (iii) 900 - 1100
cm$^{-1}$ and (iv) below 500 cm$^{-1}$. As we will discuss in the
next Section, the highest frequency modes originate from the
CH$_3$ methyl group vibrations, the strongest intensity modes in
the second range represent in-phase and out-of-phase bond
stretching C=C vibrations; the C-C-H bending vibrations are
localized in the third spectral range and the bond stretching C-Se
and Se-Se vibrations have energies lower then 500 cm$^{-1}$. The
lowest energy modes (below 200 cm$^{-1}$) correspond to the
rotational motion of TMTSF molecules. The influence of temperature
and laser power on Raman spectra in the 1300-1700 cm$^{-1}$
spectral range is illustrated in Figure 4. Figure 5 shows the full
width at half maximum (FWHM) and the frequency vs. temperature
dependencies of the C=C in phase stretching mode.

\section{Discussion}

\subsection{Normal state}

The unit cell of (TMTSF)$_2$PF$_6$ ($[(CH_3)_4C_6Se_4]_2PF_6$)
consists of one formula unit with 59 atoms in all. Since there is
a large number of atoms in the unit cell, we can expect a very
large number of optically active modes. All atoms (except P-atom
which is in (a) position-center of inversion) have 2(i) position
symmetry (C$_1$) of $P\overline{1}$ space group \cite {a4}.
Factor-group-analysis (FGA) yields:

C, Se, H, F (C$_1$): $\Gamma= 3A_g + 3A_u$

P (C$_i$): $\Gamma=3A_u$

Summarizing these representations we obtain the irreducible
representations of (TMTSF)$_2$PF$_6$ vibrational modes:

\begin{equation}
\Gamma_{P\overline{1}}= 87A_g + 90 A_u \label{1}
\end{equation}

According to this representation one can expect 177 modes from
which 87 are Raman active. Experimentally, the number of observed
modes is less then 25. The missing modes are not seen since their
intensity is probably below the noise level of our measurements.

Let us consider first the properties of each molecule in
(TMTSF)$_2$PF$_6$, separately. Three modes were observed in the
Raman spectra of hexafluorphosphate (PF$_{6}^{-}$) in different
inorganic compounds \cite {a23}. The strongest mode appears at
about 750 cm$^{-1}$ and two other modes at about 575 cm$^{-1}$
(the intensity of this mode is 12\% of the strongest mode
intensity) and at 475 cm$^{-1}$ (20\%). We did not find any
evidence of 750 cm$^{-1}$ and 575 cm$^{-1}$ PF$_{6}^{-}$ Raman
modes in  Raman spectra of (TMTSF)$_2$PF$_6$ (see Fig. 3). At
about 475 cm$^{-1}$ one mode exists, but, because we did not find
the strongest intensity mode of PF$_{6}^{-}$, we concluded that
this mode belongs to the TMTSF molecule. Thus, the Raman spectra
of (TMTSF)$_2$PF$_6$ can be considered by comparison only with
corresponding spectra of the TMTSF molecule. Vibrational
properties of the TMTSF neutral molecule and its radical cation
TMTSF$^+$ have been subject of experimental and theoretical
studies in Refs. \cite {a16, a20, a21}. We used these results to
assign our spectra, Table I. Last column in Table I shows the
assignment of the observed modes obtained using normal coordinate
analysis based on the valence-force-field (VFF) model \cite{a16}.
As it is shown in Table I, our results are in a very good
agreement with the previously published data. Beside that, we
observed 6 new modes in (TMTSF)$_2$PF$_6$ for the first time.
Because the carriers (holes) are located in the central part of
the TMTSF molecule (C$_2$Se$_4$ fragment) we have paid special
attention to spectral region below 500 cm$^{-1}$ (C-Se and Se-Se
vibration, see normal coordinates $\nu_{10}$ and $\nu_{11}$ of
these modes, Fig.1(a)) and between 1400 and 1700 cm$^{-1}$
($\nu_4$=1464 cm$^{-1}$ mode, in-phase C=C stretching and
$\nu_3$=1602 cm$^{-1}$ mode, out-of-phase C=C stretching).
Contrary to Ref. \cite {a16}, we found that $\nu_{11}$ mode at 263
cm$^{-1}$ (Se-Se stretching) exists both in the (TMTSF)$_2$PF$_6$
and in the neutral TMTSF$^0$ molecule. Also, 284 cm$^{-1}$ mode
($\nu_{10}$, Se-C stretching) appears in (TMTSF)$_2$PF$_6$ and in
an ion TMTSF$^+$ at the same frequency. It was reported in Ref.
\cite {a16} that these modes, due to ionization, have huge
positive frequency shifts (+41 cm$^{-1}$ ($\nu_{10}$), +22
cm$^{-1}$ ($\nu_{11}$)) from neutral molecule to the TMTSF$^+$
ion. We have revealed that no frequency shift exists for these two
modes, and consequently these modes show negligible coupling with
charge carriers.

A remarkable negative frequency shift is found for the C=C
vibrational modes. The $\nu_4$ mode frequency shift of about -70
cm$^{-1}$ (-140 cm$^{-1}$) is observed between the neutral
TMTSF$^{0}$ molecule and the (TMTSF)$_2$PF$_6$ (TMTSF$^+$ ion).
The $\nu_3$ mode shows also negative shift of about -23 cm$^{-1}$
between TMTSF$^0$ and (TMTSF)$_2$PF$_6$. The observed frequency
shift is a consequence of the coupling between charge, located at
the central fragment of the molecules, and the C=C stretching
A$_g$ vibrations (arising due to the electron-molecular-vibration
(EMV) coupling) \cite{a24}. As stated in \cite{a18}, these two
modes have the strongest EMV coupling constants among all observed
modes in the TMTSF salts. It is interesting to note that a
negative frequency shift of the C=C modes between the neutral
TMTSF molecule and the same molecule in (TMTSF)$_2$PF$_6$ {\it is
exactly twice less than the same frequency shift between the
neutral TMTSF molecule and its radical ion, as a consequence of
quarter-filling.}

In Fig. 4(a)-(f) we present the Raman spectra of (TMTSF)$_2$PF$_6$
measured for the (b'b') polarization at different temperatures in
the 1300 - 1700 cm$^{-1}$ range. Spectrum in Fig. 4(f) is observed
at 10 K but with an increased excitation power by a factor of 12
in comparison with the 10 K measurements shown in Fig. 4(e).
Frequency shift of this mode of about 3 cm$^{-1}$ to lower energy
and mode broadening (FWHM=7.5 cm$^{-1}$) are clear indications of
the sample heating. In order to analyze the frequency vs.
temperature dependence of high intensity modes we measured the
Raman spectra for the (b'b') polarization in the wide temperature
range with very low excitation level ($<$0.1 kW/cm$^2$). The
frequency and FWHM vs temperature dependencies of 1464 cm$^{-1}$
mode are given in Fig. 5. The same kind of temperature dependence
is found for the second C=C mode ($\nu_3$). As it is shown in Fig.
5(b) the frequency of the C=C mode has a parabolic decrease with
temperature. We found that this dependence can be described as
$\omega(cm^{-1})= 1464.3 - 5.5 \times 10^{-3}T - 48 \times
10^{-6}T^2$, where T is a temperature in K. Using these data (Fig.
5) we estimated the actual temperature of the sample in the
spectra from Fig. 4(f) as 225 K. The mode at about 1565 cm$^{-1}$
shows a stronger frequency vs. temperature dependence by lowering
temperature from 300 to 50 K than the C=C modes. Because the
frequency of this mode does not change below 50 K (see the right
panel of Fig. 4), we concluded that this mode originates from the
H-C-H bending vibrations. As proposed in Ref. \cite {a13} dealing
with the NMR data, the motion of methyl group is frozen at about
50 K, fully in an agreement with our findings.

\subsection{Spin-density-wave state}

Discussion of the magnon or spin gap excitations in
quasi-one-dimensional chains is questionable without the
theoretical calculation of the magnon dispersion. Namely, it is
well known that the non-dimerized Heisenberg chain with
antiferromagnetically coupled localized 1/2 spins does not have
spin-gap. To the best of our knowledge there are no magnon
dispersion calculations for a quarter-filled dimerized chain with
correlated electrons, the case under consideration here.

In order to calculate magnon dispersion we started from the
(TMTSF)$_{2}$X Hamiltonian \cite{a25} in the form:
\begin{eqnarray}
\cal H &=&-t_{1}\sum\limits_{i\text{ }even,l,\sigma }\left(
c_{i,l,\sigma }^{+}c_{i+1,l,\sigma }+H.c.\right)
-t_{2}\sum\limits_{i\text{ }odd,l,\sigma }\left( c_{i,l,\sigma
}^{+}c_{i+1,l,\sigma }+H.c.\right)\nonumber\\&&
-t_{b}\sum\limits_{i,l,\sigma }\left( c_{i,l,\sigma
}^{+}c_{i,l+1,\sigma }+H.c.\right) +U\sum\limits_{i,\text{
}l}c_{i,l,\uparrow }^{+}c_{i,l,\uparrow }c_{i,l,\downarrow
}^{+}c_{i,l,\downarrow }.
\end{eqnarray}

Here $t_{1,2}$ and t$_b$ are alternating hopping energies along
the chain (indices $l$ and $i$ denote chains and chain sites,
respectively), and along b-axis, respectively (Fig.1(b)) and U is
on-site Coulomb repulsion. According to the quantum chemical
calculations \cite{a26} and spectroscopy data for the TMTSF
molecule, the charge density profile of the hole is located around
the central fragment C$_{2}$Se$_{4}$. The Se $4s$ and $4p$
orbitals contribute to the conduction bands with a density such as
one hole (or three electrons) per (TMTSF)$_2$ dimer (unit cell).
Furtheron we will use the hole representation. The charge
localization at the centers of the TMTSF molecules and a small
overlap of the molecular orbitals in solid means that carriers
prefer to stay at the TMTSF sites than to move. This leads to an
enhancement of the intraTMTSF Coulomb repulsion (the
Anderson-Hubbard parameter U) of carriers and to the narrowness of
energy bands in the (TMTSF)$_{2}$PF$_{6}$ compound. For the
$\kappa$-BEDT-TTF$_2$X salts with the central fragment C$_2$S$_4$
the recent estimations \cite {a27, a28} have shown that the U
values (U=3.56 eV - 4.21 eV \cite{a28} or even U=5.37 eV
\cite{a27}) are extremely high with respect to the hopping energy
$t\sim0.1$ eV. We can assume that in the TMTSF molecule U=4 eV.
This value can be estimated from the atomic energy 1Ry
(characteristic electron correlation energy for light atoms with
size 1\AA) multiplied by the ratio of this size (around 1\AA) to
the size of the C$_{2}$Se$_{4}$ fragment (3.5\AA).

In Eq.(2), the hopping Hamiltonian is diagonalized in the standard
way and the carrier energy dispersions are given by $\varepsilon
_{p}^{\pm
}=-2t_{b}\cos p_{x}\pm \sqrt{t_{1}^{2}+t_{2}^{2}+2t_{1}t_{2}\cos p_{y}}\ $%
leading to non-correlated anisotropic band structure. The
dimerization band gap between the bonding and antibonding
non-correlated bands is $\Delta _{d}=2\left( \left|
t_{1}-t_{2}\right| -2t_{b}\right) $.

Due to the on-site electron interaction $U$, for TMTSF one can
carry out the mapping of the Hamiltonian, Eq. (2), to the
projected X-operators \cite{a29,a30} such as $c_{\sigma
}=X^{0\sigma }+\sigma X^{\overline{\sigma }2}$. The tight-binding
method for correlated electrons has been shortly presented by us
earlier for study of the transition metal oxides, high-T$_{c}$
cuprates, organic materials, spin-ladders (see \cite
{a31,a32,a33,a34,a35} and Refs. there).

In Raman scattering measurements the heating of the sample due to
laser irradiation induces thermal fluctuations which can destroy
the coherent interchain hopping amplitude and reduce the
dimensionality of (TMTSF)$_2$PF$_6$ to 1D. The temperature induced
dimensional crossover  has been revealed by transport properties
\cite {a6}, a microwave conductivity \cite{a36}  and a specific
heat \cite{a37} measurements. According to these investigations
the crossover temperature $T'$ varies in the range 3 K-35 K. The
theoretical estimations give even smaller magnitudes of $T'$. In
the (TMTSF)$_2$PF$_6$ model as weakly coupled Luttinger chains
$T'\sim(t_b/t_a^\alpha)^{1/(1-\alpha)}(t_a=t_1=t_2)$ with
$\alpha=(K_\rho+1/K_\rho-2)/4$ \cite{a38}, and values of $K_\rho$
in the range 0.2-0.25 or 0.35 \cite{a39}. In contrast, the recent
Hall effect and electrical conductivity measurements \cite{a6} in
(TMTSF)$_2$PF$_6$ confirm the Fermi-liquid description at
$T\lesssim300 K$ with the possible 1D Luttinger liquid picture
well above room temperature. At the assumption that in our
experiment we are dealing with $T>T'$, in the range of temperature
reduced dimensionality, the minimal model for (TMTSF)$_2$PF$_6$
salt is the dimerized quarter-filled linear chain with correlated
electrons ($t_b$=0).

To calculate the magnon energy dispersions in SDW-state in
TMTSF-chain, first of all we split the lattice into two
sublattices, {\it a} and {\it b}, with opposite spins. In
sublattice a(b) with spin projection up (down) the localized
magnon (zeroth order approximation) has energy $\omega
_{+-}=\varepsilon _{-}-\varepsilon
_{+}=2H$ $\left( \omega _{-+}=\varepsilon _{+}-\varepsilon _{-}=-2H,\text{%
the Bohr magneton }\mu _{B}=1\right) $ in magnetic field $H$. In
the second order of the perturbation theory the inverse magnon
Green's function is presented by the two-by-two matrix as follows:

\begin{equation}
D_{+-}^{-1}\left( p,\omega \right) =
\begin{array}{c}
+- \\ -+
\end{array}
\left(
\begin{array}{cc}
\frac{-i\omega}{f_{+-}^a} + 2H +\Pi _{aa} & \Pi _{ab} \\ \Pi _{ba}
& \frac{i\omega}{f_{-+}^b} - 2H +\Pi _{bb}
\end{array}
\right) .
\end{equation}

In Eq.(3) the diagonal, $\Pi _{aa,bb}$, and non-diagonal, $\Pi
_{ab,ba}$, magnon self-energies (polarization operators) are
defined by the sum of diagrams presented in Figure 6. For a
quarter filled chain the magnon correlation factors are defined as
$f_{\sigma \overline{\sigma
}}=n_{\sigma}-n_{\overline{\sigma}}=n_{\sigma}=\frac{1}{2}$. Each
polarization diagram includes a pair of hole hops between
sublattices. The internal solid lines are the hole Green's
functions $G_{\alpha}^{a,b}(q,\omega)=1/(-i\omega_n+\xi_q-\mu)$,
with correlated energies
$\xi_q=\frac{U}{2}\pm\frac{1}{2}\sqrt{U^2+4f_a^{0+}f_b^{2-}\varepsilon_q^2}$,
where $\varepsilon _q$ is the non-correlated tight-binding
dispersion, $\mu$ is the chemical potential and $f_{a,b}^{0+,2-}$
are the fermionic correlation factors (Fig.6). In paramagnetic
phase the correlation factors are governed by a carrier
concentration. In each sublattice the spin ordered electrons can
be considered as spinless fermions, for which the fermionic
correlation factors $f_{a,b}^\alpha=1$

The diagonal magnon-self energies (Fig. 6(a)) are momenta
independent and have analytical expressions such as
\begin{equation}
\Pi _{aa,bb}=-4\frac{t_1^2+t_2^2}{U}.
\end{equation}
For small difference between hopping integrals along slightly
dimerized chains, $|t_1-t_2|<<U$, of the (TMTSF)$_2$PF$_6$ salt
(the existing evaluations for hopping energies are $t_1$=147 meV
\cite{a40}, 252 meV \cite{a41} and $t_2$=144 meV \cite{a40}, 209
meV \cite{a41}) the non-diagonal self-energies, $\Pi _{ab}$ and
$\Pi _{ba}$ are momenta dependent and composed from diagrams in
Figs. 6(b), 6(c), which can be expressed analytically as follows
\begin{equation}
\Pi
_{ab}=\frac{4}{U}(t_1^2+t_2^2e^{2ip})+o(\frac{t_1^2-t_2^2}{U}),\ \
\ \Pi
_{ba}=\frac{4}{U}(t_1^2+t_2^2e^{-2ip})+o(\frac{t_1^2-t_2^2}{U}).
\end{equation}

After an analytical continuation, $i\omega \rightarrow \omega
+i\delta $ and putting $H\rightarrow0$, the zeros of the inverse
magnon Green's function, Eq. (3), (poles of the Green's function
$D_{+-}\left( p,\omega \right) $) give the spin-wave spectrum

\begin{equation}
\omega(p)= 2n_{\sigma}\frac{4t_1t_2}{U}\sin
p=\frac{4t_1t_2}{U}\sin p,
\end{equation}

for a considered model of the (TMTSF)$_{2}$PF$_{6}$ salts with one
carrier per dimer (TMTSF)$_2$ ($n=1$, $n_{\sigma}=1/2$).

It is important to note that in the limiting case of a half-filled
nondimerized chain ($n$=2, $n_{\sigma}$=1, $t_1=t_2=t$) the Eq.(6)
gives the conventional gapless spectrum of spin-waves as $\omega
\left( p\right) =$ $2J\sin p$ with exchange parameter
$J=4t^{2}/U$. In this case the magnon amplitude is higher {\it by
factor 4/$\pi$ only}, than exact the des Clozeaux-Pearson result
for spins 1/2 in the Heisenberg chain, $\frac{\pi}{2}J$sin$p$
\cite {a42,a43}.

In Raman spectra at 4 K we found no difference with the
corresponding spectra above 10-12 K. This has to be expected
because:

{\it (i)} It was reported \cite {a44} that the spin-lattice
coupling in the SDW state is very weak in (TMTSF)$_{2}$PF$_{6}$,
and consequently the phonon frequency shift at SDW transition is
not observable. Also, the heat capacity does not change
significantly at T$_{SDW}$ \cite{a45}.

{\it (ii)} Raman scattering due to magnetic excitation is usually
very weak in low-dimensional systems. Because of a weak
antiferromagnetic ordering in (TMTSF)$_2$PF$_6$ it is hard to
expect an appearance of new spin-related modes. Effect of magnetic
fluctuations on Raman scattering was analyzed only in the
$\kappa$-BEDT-TTF$_2$Cu(NCS)$_2$ organic superconductor
\cite{a19}. It was shown that the frequencies of two modes soften
in the temperature range where antiferromagnetic spin fluctuations
(80 K) have been observed. In our case the critical temperature
(12 K) is too close to our experimental limit (4K) and no
frequency shift is observed.

{\it (iii)} For TMTSF salts Yamaji predicted \cite{a46} that a
phase transition from metallic to SDW state occurs in two steps.
At T$_{SDW}$, there is a transition, first from metallic into
semimetallic SDW state and after that at T$^*<$T$_{SDW}$ into an
insulating SDW state. Such picture in (TMTSF)$_2$PF$_6$ is
verified by transport measurements \cite {a47,a48,a49} under
pressure and/or in high magnetic fields  \cite {a50, a51, a52}.
Furthermore, the proton NMR measurements \cite {a53} have shown no
nuclei damping below T$^*\simeq$3.5 K and the Korringa-like
$^{77}Se$ spin lattice relaxation rate has been found between
T$^*$ and T$_{SDW}$ \cite {a54}, with activated character below
T$^*\simeq$4 K. Thus, the coexistence of metallic and SDW states
above T$^*$ leads to an incomplete SDW gap and also hinders an
observation of the spin modes in Raman spectra.

{\it (iv)} Finally, we can not totally exclude the effect of the
sample heating by laser irradiation, which can destroy the
coherent interchain hopping amplitude and reduce the
dimensionality of (TMTSF)$_2$PF$_6$ to 1D.

In conclusion, we have measured the polarized Raman spectra of
(TMTSF)$_2$PF$_6$ single crystals in a wide spectral and
temperature range using different laser energies. We observed and
assigned 23 Raman active modes, of which 6 are observed for the
first time. We found that Se-Se and C-Se stretching modes at 263
cm$^{-1}$ and 284 cm$^{-1}$, respectively do not have a positive
frequency shift in comparison with corresponding modes of (TMTSF)
neutral and ionized molecule, i.e. these modes are not coupled
with charge carriers. Only the C=C stretching modes at about 1464
and 1602 cm$^{-1}$ show the strong electron-molecular-vibration
coupling. We found that the frequency vs temperature dependence of
the strongest intensity mode is $\omega(cm^{-1})= 1464.3 - 5.5
\times 10^{-3}T - 48 \times 10^{-6}T^2$, where T is a temperature
in K. The mode at about 1565 cm$^{-1}$ shows no temperature shift
below 50 K due to the CH$_3$ methyl group motion freezing.

We calculated magnon dispersion for the one-dimensional correlated
quarter-filled chains for the first time. Our calculation for a
temperature reduced 1D and slightly dimerized (TMTSF)$_2$PF$_6$
salt in the SDW phase does not show the spin gap at
quarter-filling. We found no difference in the Raman spectra
between the normal and the SDW state as a consequence of a very
weak antiferromagnetic ordering in the SDW phase or possibly due
to an imperfect SDW phase in the temperature range 4 K$<T<$12 K.

\section*{Acknowledgment}

Z.V.P., V.A.I and O.P.K acknowledge support from the Research
Council of the K.U.  Leuven and DWTC.  The work is supported by
the ESF Program FERLIN and the Belgian IUAP and FWO-Vl and GOA
Programs. V.A.I. acknowledges illuminating discussions with K.
Yamaji about the (TMTCF)$_2$X phase diagram and with Yu. Kagan.

\clearpage

\begin{figure}
\caption {Schematic representation of the (TMTSF)$_2$PF$_6$
 crystal structure in
the (a) (010) and (b) (001) plane. The dashed lines represent the
unit cell. } \label{fig1}
\end{figure}

\begin{figure}
\caption {Raman scattering spectra of  (TMTSF)$_2$PF$_6$ for the
(aa) polarization at 10 K.} \label{fig2}
\end{figure}

\begin{figure}
\caption {Raman scattering spectra of  (TMTSF)$_2$PF$_6$ for the
(b'b') polarization at 10 K.} \label{fig3}
\end{figure}

\begin{figure}
\caption {Raman scattering spectra of  (TMTSF)$_2$PF$_6$ for the
(b'b') polarization at different temperatures and laser power:
(a)-(e) P=0.1 kW/cm$^2$, (f) P=1.2 kW/cm$^2$. } \label{fig4}
\end{figure}

\begin{figure}
\caption {Frequency and FWHM vs temperature dependencies of the
1464 cm$^{-1}$ mode. The solid line is parabolic fit
$\omega(cm^{-1})= 1464.3 - 5.5 \times 10^{-3}T - 48 \times
10^{-6}T^2$. } \label{fig5}
\end{figure}

\begin{figure}
\caption {The second order polarization operator diagrams for
"diagonal", $\Pi _{aa,bb}$, and "non-diagonal", $\Pi _{ab,ba}$,
magnon self-energies in Eq.(3). Solid lines denote magnon ($+-,
-+$) and fermion ($\alpha, \beta$) Green's functions, whereas
dashed lines ("interactions") mark the hole hopping energies
between sublattices of dimerized TMTSF-chain. The "triangular"
arrows stand for the on-site transitions labeled by $su(2,2)$
superalgebra roots $\alpha, \beta$. The circles stand for
fermionic correlation factors $f^{pq}=\langle
X^{pp}+X^{qq}\rangle_0$. (a) For $\Pi _{aa}$ (left diagram) the
on-site transitions are labeled by roots $\alpha=2- (0-)$ and
$\beta=+0 (+2)$. For $\Pi _{bb}$(right diagram) the on-site
transitions are labeled by roots $\alpha=0+ (2+)$ and $\beta=-2
(-0)$. (b) Polarization diagrams for the self energy $\Pi
_{ab}(p)$. (c) Polarization diagrams for the self-energy $\Pi
_{ba}(p)$.} \label{fig6}
\end{figure}

\clearpage
\renewcommand{\arraystretch}{0.8}

\begin{table}
\caption{Raman mode frequencies of TMTSF$_2$PF$_6$, TMTSF neutral
molecule and TMTSF$^+$. $A_g$ are Raman active modes of the TMTSF
molecule within $D_{2h}$ symmetry.}
\begin{tabular}{cccccccc}
\hline \hline

TMTSF$_2$PF$_6$ & TMTSF$_2$PF$_6$ &TMTSF$_2$PF$_6$& TMTSF$^{0}$ &
TMTSF$^+$ &&& Assignment
\\
This work& Ref.[20]&Ref.[21]& Ref.[20] & Ref.[16] &$A_g$& &
Ref.[16]\\ \hline 156 & - & 155&173 & - & $\nu_{12}$&&\\ 181 & - &
&-&180 & - & &  \\ 264 & - &263& 263 &-& $\nu_{11}$& & CH$_3$ -
bending
\\ 284 & 278 & 281&276 & 285 & $\nu_{10}$& & Se-Se stretching
\\ -& - & - &-& 317 & & &
-
\\ 331 & - & 328& 328 & - & -&&\\ 397 & - &398& - & - & -&&\\ 452 & - &452& 453 & 452 &
$\nu_{9}$&& Se-C stretching
\\475 & 478 &-& 472 & - & -&&\\ - & -&- & 603 & - &&&\\- & 675&- & 682 & - &
-&&\\916 & 917 & 918&916 & 924 & $\nu_{8}$&&\\ 921 &- &-& - &
-&&&\\ 936 & - &936&- & - & &\Bigg \}&C-C-H bending\\- & 1010&- &
1018 & - &&&\\1071 & - &1070& - & - &$\nu_{7}$&&\\1092 & - & -&- &
- &&&\\- & - & -&1167 & -&&&\\- & -&- & 1225 & 1245 &&&\\ 1388 &
1381 &-& 1384 & 1386 & $\nu_{6}$&\Bigg \}&H-C-H bending\\1452 & -
& 1452&1444 & - & $\nu_{5}$&&\\- & -&- & 1503 &- &&&\\ - & - &-&
1530 & - &&&\\ 1464 & 1463& 1464& 1539 & 1399 & $\nu_{4}$&& C=C in
phase stretching
\\1477 & - & 1479&- & - &\\ 1521 & 1547 & - & - &&&\\ 1565 & 1559 & 1565&1589
& - &&\Bigg \}&H-C-H- bending\\1582 & - & 1582&- & - &&&\\ 1602 &
- & 1601&1625 & 1573 & $\nu_{3}$&& C=C out of phase
stretching\\2800 &-&-&-&-&$\nu_{2}$&&C-H stretching
\\2908&2906&-&-&-&$\nu_1$&&C-H stretching\\
 \hline\hline
\end{tabular}
\end{table}


\clearpage



\begin{thebibliography}{99}



\bibitem{a1} D. J\'erome, A. Mazaud, M. Ribault, and K. Bechgaard, J.
Physique Lett., {\bf41}, L95 (1980).
\bibitem{a2} T. Ishiguro, K. Yamaji, and G.
Saito,{\it Organic Superconductors, }Springer Series in
Solid-State Sciences, Vol. 88 (Springer, Berlin, 1998) and Refs.
there.
\bibitem{a3}L. P. Gor'kov, J. Phys. I (France), {\bf6}, 1697
(1996).
\bibitem{a4}N. Thorup, G. Rindorf, H. Soling, and K. Bechgaard,
Acta Crys. B {\b 37}, 1236 (1981);
\bibitem{a5}K. Bechgaard, M. M.
Nielsen and F. C. Krebs, J. Phys. IV (France)10, Pr3-11, (2000).


\bibitem{a6} G. Mih\'aly, I. K\'ezsm\'arki, F. Z\'amborszky and L. Forr\'o, Phys. Rev. Lett.
{\bf 84} 2670 (2000).
\bibitem{a7} J. Moser, J. R. Cooper, D. J\'erome, B. Alavi, S. E.
Brown and K. Bechgaard, Phys. Rev. Lett. {\bf84}, 2674 (2000).

\bibitem{a8} M. Dressel, Physica C {\bf317-318},  89 (1999).
\bibitem{a9}L. Degiorgi, M. Dressel, A. Schwartz, B. Alavi and G.
Gr\"uner, Phys. Rev. Lett. {\bf76}, 3838 (1996).
\bibitem{a10} V. Vescoli, L. Degiorgi, W. Henderson, G. Gr\"uner, K. P.
Starkey, and L. K. Montgomery, Science {\bf281}, 1181 (1998).

\bibitem{a11} S. Zherlitsyn, G. Bruls, A. Goltsev, B. Alavi and M.
Dressel, Phys. Rev. B {\bf59}, 13861 (1999).

\bibitem{a12}W. M. Walsh Jr., F. Wudl, G. A. Thomas, D. Nelewajek,
J. J. Hauser, P. A. Lee, and T. Poehler, Phys. Rev. Lett. {\bf
45}, 829 (1980).
\bibitem{a13}J. C. Scott, H. J. Pedersen, and K. Bechgaard, Phys.
Rev. B {\bf24}, 475 (1981).

\bibitem{a14} K. Ichimura, O. Abe, K. Nomura, S. Takasaki, J.
Yamada, S. Nakatuiji, and H. Aazai, Synth. Metals, 103, 2097
(1999).
\bibitem{a15} K. Kornelsen, J.E. Ebbridge, G.S.Bates, Phys. Rev. B
{\bf55}, 9162 (1987).

\bibitem{a16}M. Meneghetti, R. Bozio, I. Zanon, C. Pecile, and C.
Ricotta, J. Chem. Phys. {\bf80}, 6210 (1984).
\bibitem{a17}S. Sugai, H. Mori, H. Yamochi and G. Saito, Phys. Rev.
B {\bf47}, 14374 (1993).


\bibitem{a18}D. Pedron, G. Visentini, E. Cecchetto, R. Bazio, J. M.
Williams and A. Schlueter, Synth. Metals, {\bf85}, 1509, (1997).
\bibitem{a19} Y. Lin, J. E. Eldridge, H. H. Wang, A. M. Kini, M. E.
Kelly, J. M. Williams and J. Schlueter, Phys. Rev. B {\bf58}, R599
(1998).
\bibitem{a20}K. Iwahana, H. Kuzmany, F. Wudl, E. Aharon-Shalom,
Mol. Cryst. Liq. Cryst., {\bf79}, 39 (1982).
\bibitem{a21} M. Krauzman, H. Poulet, and R. M. Pick, Phys. Rev. B
{\bf33}, 99 (1986).


\bibitem{a22} Ref\cite{a2}, p.448
\bibitem{a23} R. A. Niquist, C. L. Putzig, M. A. Leugers, {\it The
Handbook of Infrared and Raman Spectra of Inorganic Compounds and
Organic Salts}, Academic Press, New York, 1996, p.19.



\bibitem{a24} M. J. Rice, Solid State Commun., {\bf 31}, 93
(1979).
\bibitem{a25} V. A. Ivanov and K. Yamaji, unpublished.

\bibitem{a26} R. V. Metzger, J. Chem. Phys. {\bf75}, 482 (1981).

\bibitem{a27} Y. Okuno and H. Fukutome, Solid State Commun. {\bf 101}, 355 (1997).

\bibitem {a28}A. Fortunelli and A. Painelli, Phys. Rev. B {\bf55}, 16088
(1997).
\bibitem {a29} S. Okubo, Prog. Theor. Phys., {\bf27}, 949 (1962).

\bibitem{a30} J. Hubbard, Proc Roy. Soc. A
{\bf276}, 238 (1963), {\it ibid.}  {\bf277}, 237 (1964).

\bibitem{a31} V. A. Ivanov, Physica B {\bf186-188},
921 (1993).
\bibitem{a32} V. A. Ivanov, J. Phys. Condens. Matter, {\bf6},
2065(1994); Physica C, {\bf185-189}, 1635 (1991).
\bibitem{a33} V. A. Ivanov, E. A. Ugolkova, and M. Ye. Zhuravlev, Sov. Phys. JETP {\bf86}, 395 (1998).

\bibitem{a34}Z. V. Popovi\'c, M. J. Konstantinovi\'c, V. A. Ivanov, O. P. Khuong, R. Gaji\'c,
A. Vietkin and V. V. Moshchalkov, Phys. Rev. B {\bf62}, 4963
(2000).
\bibitem{a35} V. A. Ivanov, Z. V. Popovi\'c, O. P. Khuong, and V. V.
Moshchalkov, cond-mat/9909046.
\bibitem{a36} P. Fertey, M. Poirier, and P. Batial, Synth. Metals,
{\bf103}, 2076 (1999).
\bibitem{a37} J. C. Lazjaunias, K. Biljakovi\'c, H. Yang, and P.
Monceau, Synth. Metals, {\bf103}, 2130 (1999).
\bibitem{a38} C. Bourbonnais, J. Phys. IV (France), {\bf10}, 3 (2000).
\bibitem{a39} A. Georges, T. Giamazcki and N. Sandler,
cond-mat/0001063.

\bibitem{a40} P. M. Grant, Phys. Rev. B {\bf 26}, 6888 (1982).
\bibitem{a41} L. Ducasse, M. Abderrabba, J. Hoarau, M. Pesquer, B. Gallois and J. Gaultier, J. Phys. C {\bf19}, 3805 (1989).

\bibitem{a42} J. des Clozeaux and J. J. Pearson, Phys. Rev.
{\bf128}, 2131 (1962).
\bibitem{a43} A. A. Ovchinnkov, Sov. Phys. JETP, {\bf29}, 727 (1969).


\bibitem{a44}G. Kreuzet, C. Gaonach and B. Hamzi\'c, Synth. Metals, {\bf19}, 245 (1987).
\bibitem{a45}J. Coroneus, B. Alavi, S. E. Braun, Phys. Rev. Lett.,
{\bf70}, 2332 (1993).
\bibitem{a46}K. Yamaji, J. Phys. Soc. Japan, {\bf51}, 2787 (1982); also Ref.[2], p.102.

\bibitem{a47}J. P. Ulmet, P. Auban, A. Khmou, S. Askenazy, and A. Moradpour,
J. Phys. Lett.(Paris), {\bf46}, 535 (1985).
\bibitem{a48}S. Uji, J. S. Brooks, M. Chaparala, S. Takasaki, J.
Yamada and H. Anzai, Phys. Rev. B {\bf55}, 12446 (1997).
\bibitem{a49} J. S. Brooks, J. O'Brien, R. P. Starrett, R. G.
Clark, R. H. McKenzie, S. Y. Han, J. S. Qualls, S. Takasaki, J.
Yamada, H. Anzai, and L. K. Montgomery, Phys. Rev. B {\bf59}, 2604
(1999).

\bibitem{a50}A. Audouard and S. Ashkenazy, Phys. Rev. B {\bf52},
700 (1995).
\bibitem{a51}N. Bi\u{s}kup, S. Tomi\'c, D. J\'erome, Phys. Rev. B
{\bf51}, 11972 (1995).
\bibitem{a52}K. Maki, Phys. Rev. B {\bf47}, 11506 (1993).


\bibitem{a53}T. Takahashi, T. Harada, Y. Kobayashi, K. Kanoda,
K. Suzuki, K. Murata, and G. Saito, Synth. Metals, {\bf41-43}, 107
(1985).



\bibitem{a54}S. Valfells, R. Kuhns, A. Kleinhammes, J. S. Brooks,
W. Moulton, S. Takasaki, J. Yamada, and H. Anzai, Phys. Rev. B
{\bf56}, 2585 (1997).






\end{thebibliography}
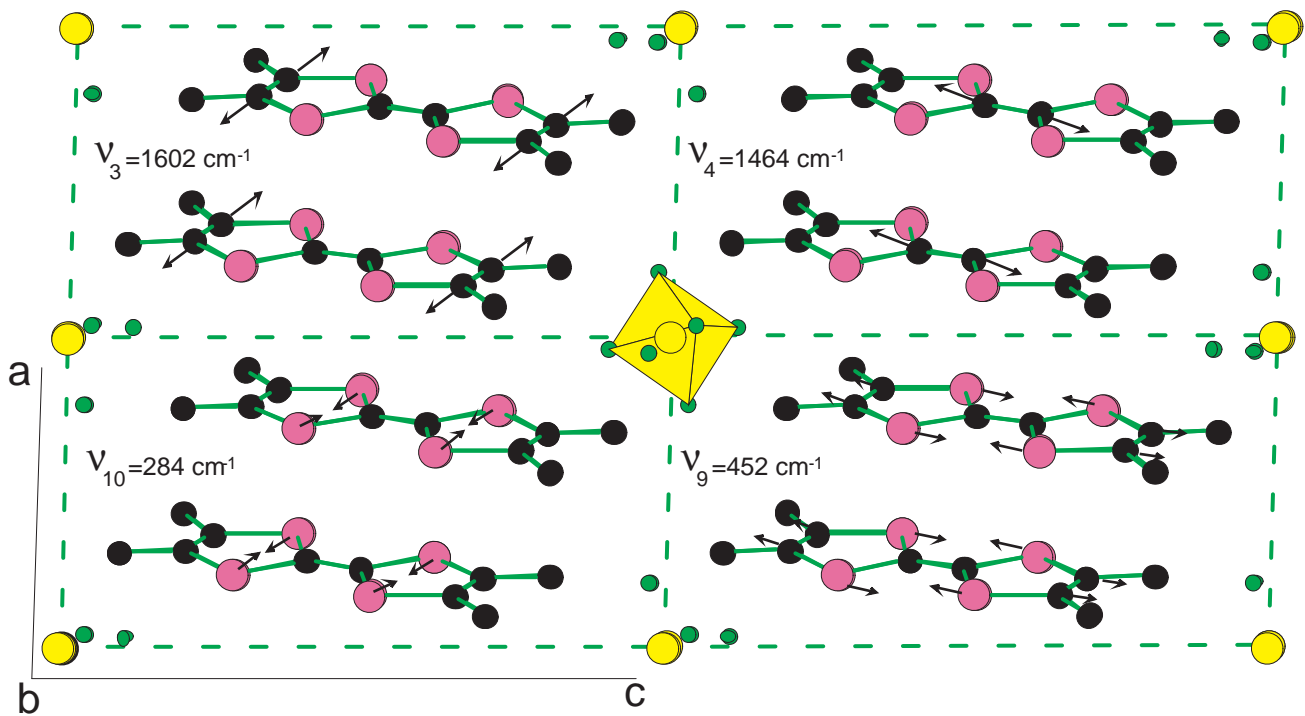
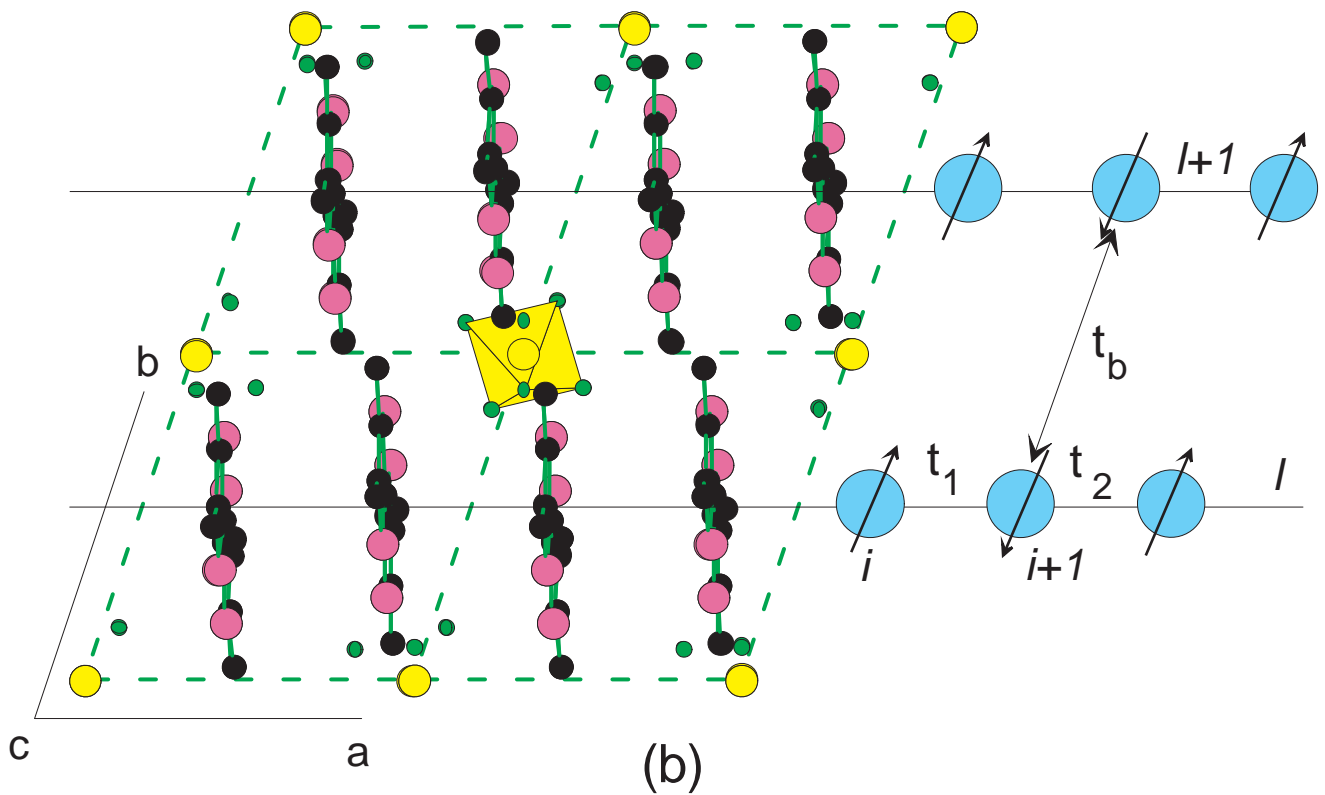
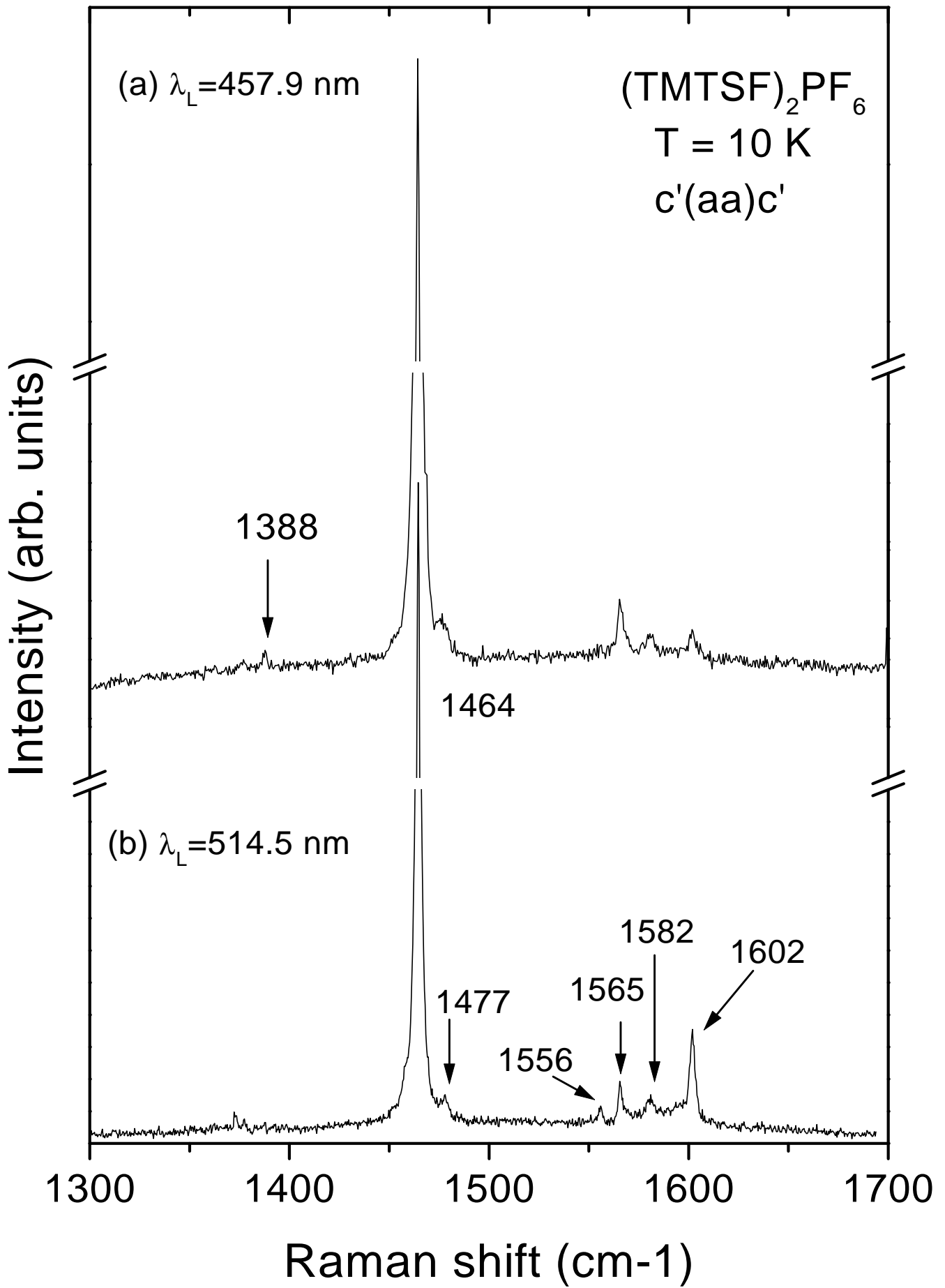
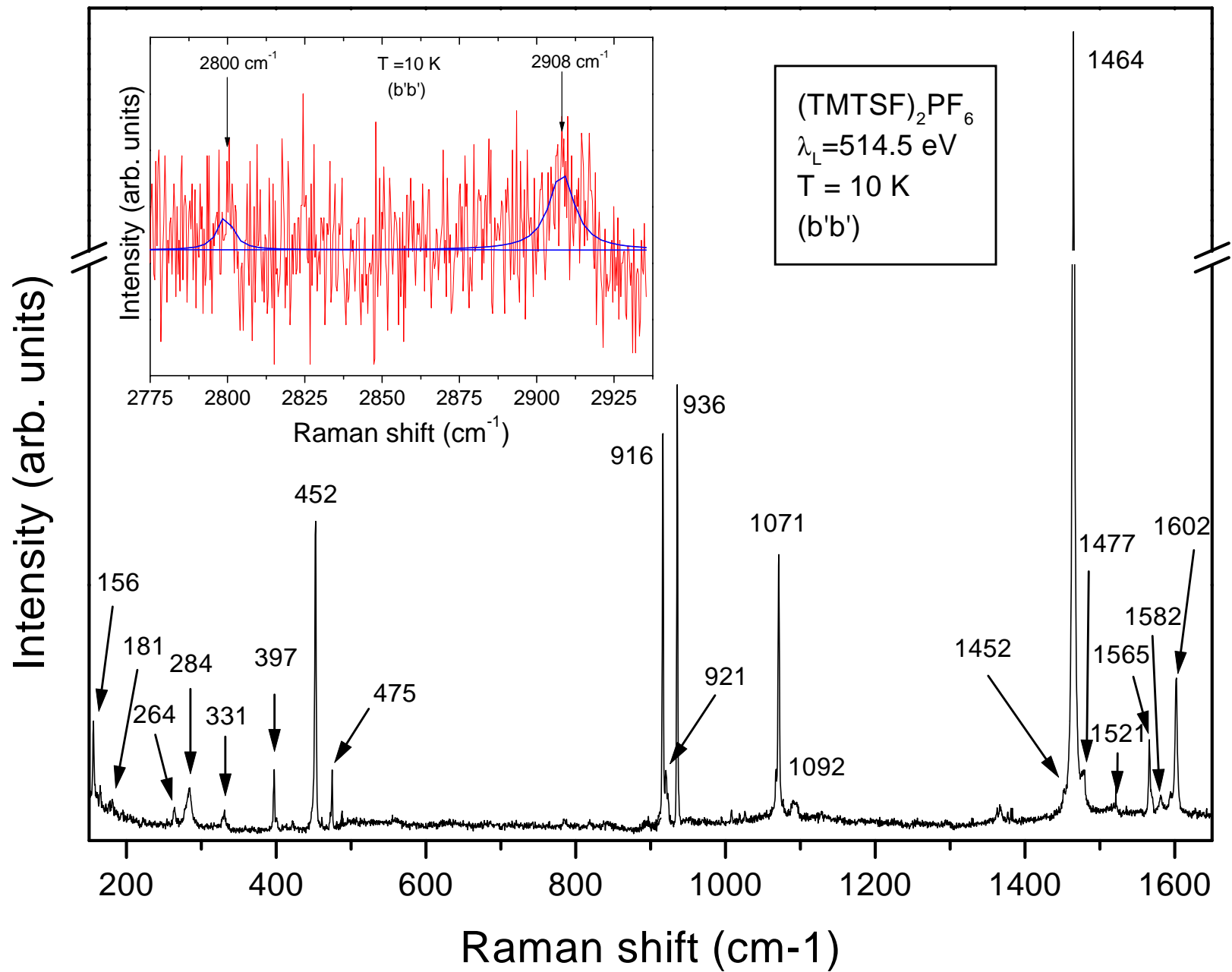
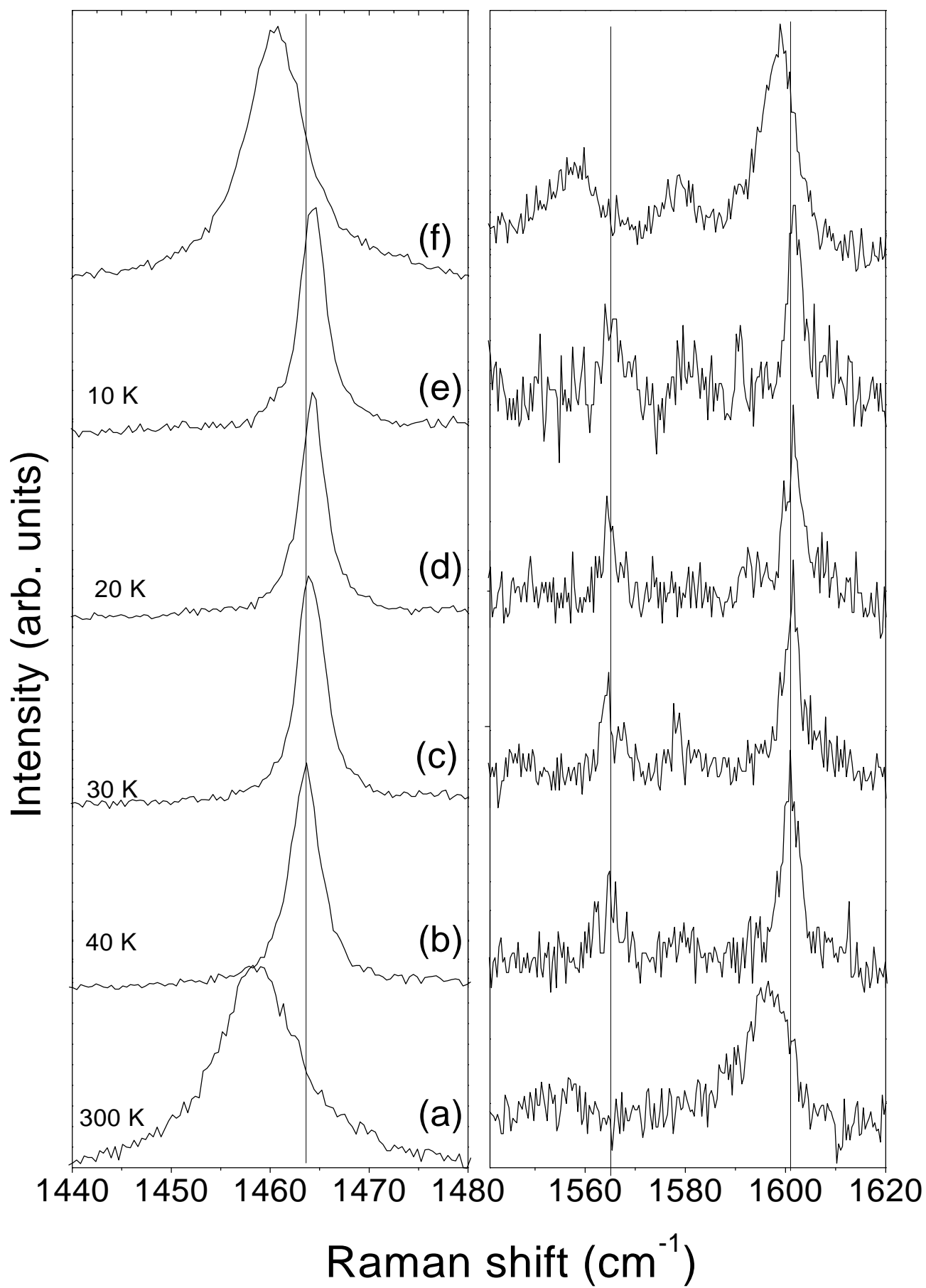
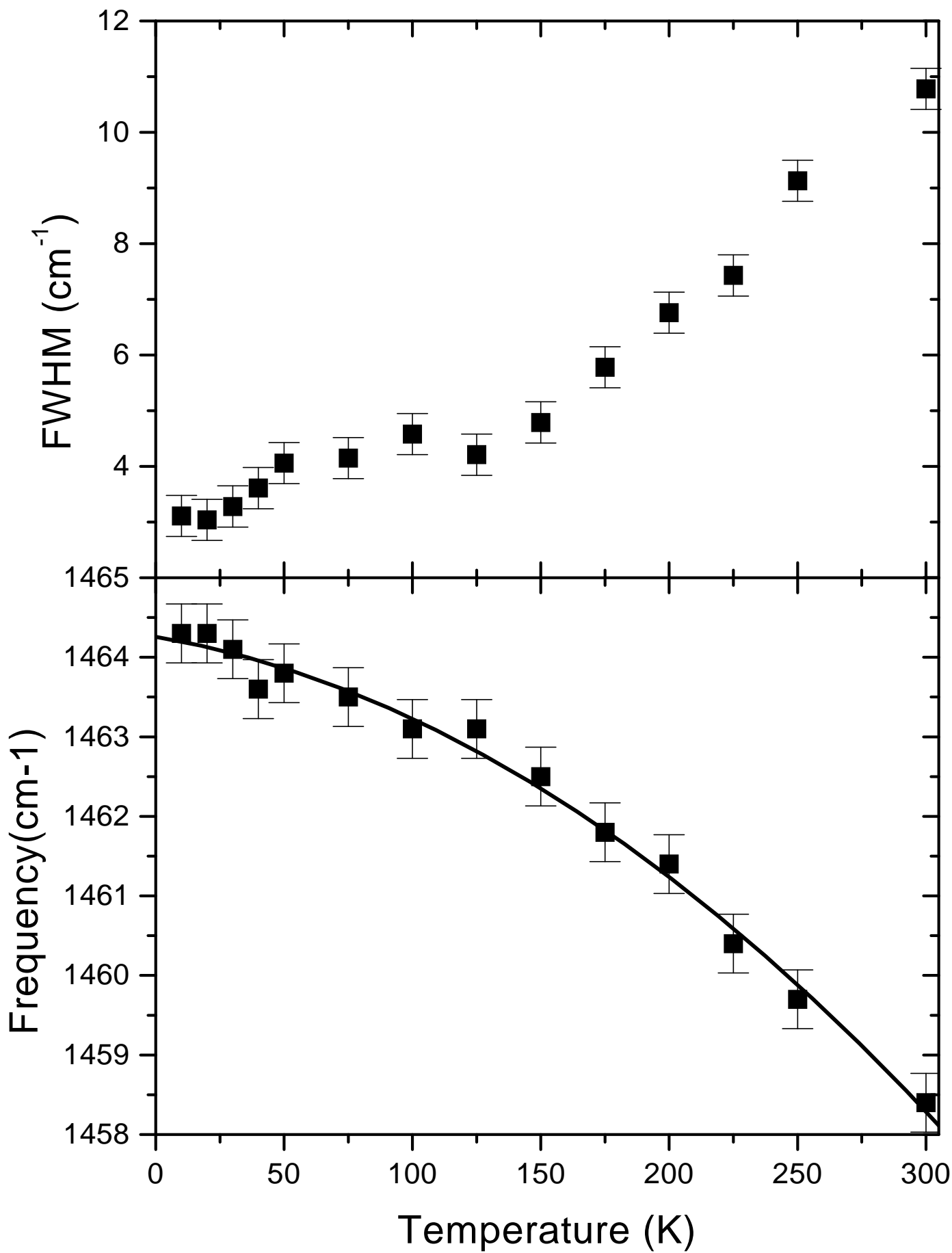
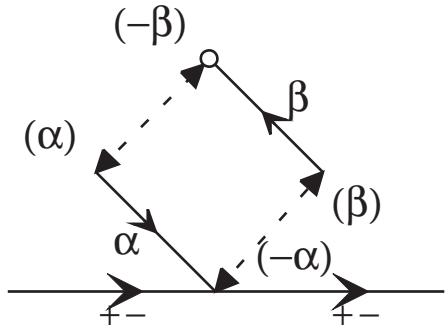
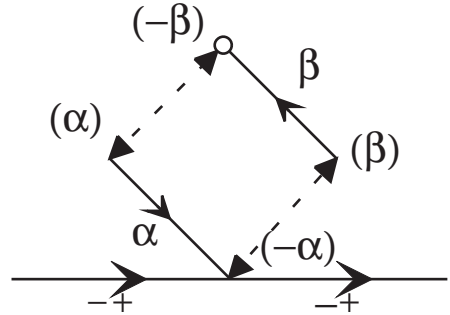
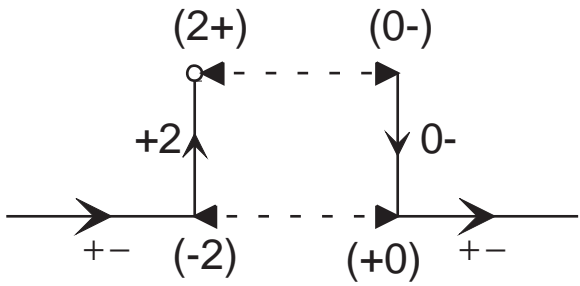
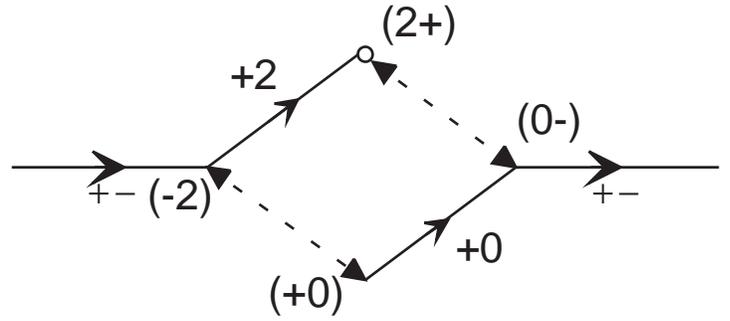
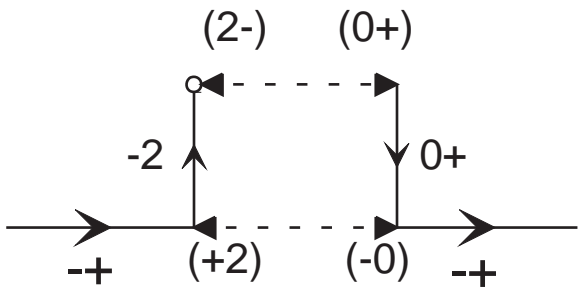
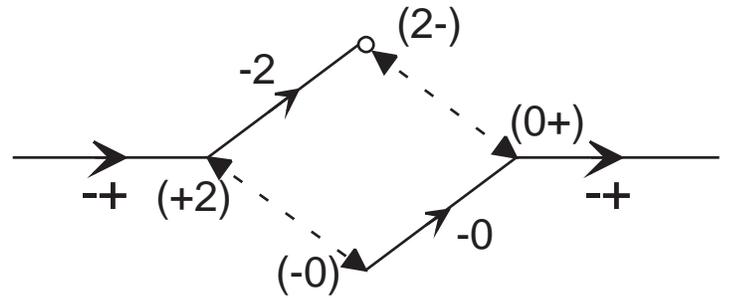
\end{document}